\documentclass[preprints,article,accept,pdftex,moreauthors]{Definitions/mdpi} 
\firstpage{1} 
\pubvolume{1}
\issuenum{1}
\articlenumber{0}
\pubyear{2025}
\copyrightyear{2025}
\datereceived{ } 
\daterevised{ } 
\dateaccepted{ } 
\datepublished{ } 
\hreflink{https://doi.org/} 

\Title{A Real-time 3D Desktop Display}

\TitleCitation{A Real-time 3D Desktop Display}

\Author{Livio Tenze and Enrique Canessa $^{*}$\orcidB{}}

\AuthorNames{Livio Tenze and Enrique Canessa}

\AuthorCitation{Tenze, L.; Canessa, E.;\\}

\address[1]{%
SciFabLab, International Centre for Theoretical Physics, Trieste, Italy
}

\corres{Correspondence: canessae@ictp.it}

\abstract{
A new extended version of the altiro3D C++ Library --initially developed to get 
glass-free holographic displays starting from 2D images-- is here introduced aiming 
to deal with 3D video streams from either 2D webcam images or flat video files. These 
streams are processed in real-time to synthesize light-fields (in Native format) and 
feed realistic 3D experiences. The core function needed to recreate multiviews 
consists on the use of MiDaS Convolutional Neural Network (CNN), which allows to 
extract a depth map from a single 2D image. Artificial Intelligence (AI) computing 
techniques are applied to improve the overall performance of the extended altiro3D 
Library. Thus, altiro3D can now treat standard images, video streams or screen 
portions of a Desktop where other apps may be also running (like web browsers, 
video chats, etc) and render them into 3D. To achieve the latter, a screen region 
need to be selected in order to feed the output directly into a light-field 3D 
device such as Looking Glass (LG) Portrait. In order to simplify the acquisition 
of a Desktop screen area by the user, a multi-platform Graphical User Interface 
has been also implemented.
Sources available at: https://github.com/canessae/altiro3D/releases/tag/2.0.0
}


\begin{document}

\section{Introduction}

Most of the popular PC-based video games and immersive displays for Virtual and Augmented 
Reality systems to render interactive true 3D visualizations, require the user to wear 
a ad-hoc device and provide solitary 3D experiences. On the other hand, there are other 
classes of glasses-free 3D screen technologies available to explore realistic results 
on-the-fly. For example, there exist large 3D screens having combinations of sensors 
to track eye movements and gaze their direction in a visual 3D environment \cite{Ban22}. 
3D Monitors utilize techniques like infrared (IR) light reflection and cameras to capture 
eye data, which are then processed to determine a specific user's viewpoint. Although such 
devices for 3D vision may lead to an impressive illusion of real-world, these still demands 
high computational processing making them ill-suited to automatically generate live streaming 
applications.

3D streaming in real time is demanding because of the huge amount of information contained 
in the light-fields to be streamed live, as compared to sending monocular frames via, e.g., 
standard 2D video streams. However, most recent alternative methods to render 3D computer 
reconstructions faster are those algorithms based on the generation of depth maps 
from monocular images for modeling the 3D visual world by Convolutional Neural Network (CNN) 
models such as MiDaS trained over large-scale RGB data-sets \cite{Ten24}. 
Starting from just a single RGB image or frame, given as the input, tremendous progress 
has been done with this alternative framework for fast 2D-to-3D image conversions. To 
represent reality starting from a given RGB image, the recent altiro3D C++ Library \cite{Ten24} 
can synthesize automatically N-number of digital images and add them sequentially into a Quilt 
collage by applying MiDaS-CNN to obtain the monocular depth estimation \cite{midas}. View 
synthesis from a single image are obtained through simple OpenCV and Telea in-painting 
techniques to map all pixels along N-viewpoints. altiro3D implements a unique pixel- and 
device-based Lookup Table (LUT) to optimize computing time \cite{Can20}. In the absence of 
this LUT procedure, it would become difficult to generate light-field images or videos in 
real-time.

The multiple viewpoints and video generated from a single image or frame can be displayed 
in modern free-view displays such as Light Field Displays (LFD) \cite{LG25}. To this end, 
it is tempting to then study the possibility of extending these 3D representations from 
single images to a more dynamical, live 3D vision. Any N-view synthesis algorithm in this 
case would require a fast conversion of each video frame into a Native light-field image
(and corresponding Quilt), in order to achieve a large-viewpoint --at a frame rate of at 
least 10 fps in order to achieve live video streaming. In this study, we made an effort to 
contribute to this field by implementing AI-generated multi-views for the display of a Desktop 
in 3D. We introduce an extended version of the altiro3D C++ Library in order to generate 
3D video streams in real-time by recreating multiple light-fields for each video frame. 
We discuss how the present extended altiro3D algorithm can create on-the-fly, 3D versions 
of 2D video streams --being displayed using any standard video player-- or any segment of 
the Desktop screen in use in which other apps may be also running. All of these streamed 
apps can be directly feed into a lenticular-based LFD in real-time. 

\section{Briefing on 3D displays}

Conventional 2.5D desktop interfaces --with basic elements in the plane "xy" such as 
apps icons and pop-up and pull-down windows menus, and the superposition of “background” 
windows in the pseudo-plane "z"-- still occupy the mainstream despite their limitations. 
Typically, with current desktop extensions it is not always possible to visualize 
and manipulate data that defines objects in a 3D space efficiently. Any software 
design and implementation of a real-time GUI as a front-end virtual tool should display 
a realistic 3D environment. In particular, it needs to be the front-end for a practical 
application with at least the same level of usability as the standard 2D interface which 
users are accustomed. These facts raise many challenges to developers.

The past two decades has witnessed important advances in the study of 3D displays. 
One of the first 3D computer interfaces introduced a set of windows arranged in a 
perspective projection (within a box) being part of the screen as the front-end 
for the user. These desktop 3D interfaces are not 3D themselves, they only allow 
users to manage files and applications in a spatial context in which a user may become 
easily disoriented. Alternative auto-stereoscopic 3D display systems may require 
special 3D glasses that are unsuitable in some environments and can cause visual 
fatigue because the fixed viewing distance and position needed for their implementation.
Parallax barrier 3D technology for 3D displays, on the other hand, reduces the images 
brightness. For an overview, and a comparative analysis, on different types of 3D user 
interfaces see Refs.\cite{Bow08,Pan21}.

Volumetric 3D display with mechanical rotating components is another kind of 3D display 
with the inherent limitation of reducing image resolution.  Computer- or Laser-generated 
holograms can reconstruct the wave-fronts of scattered light by a static object and thus
reproduce the sensation of viewing real (but small) objects. However, the difficulty to 
display color images and to generate these dynamically are main shortcomings of these 
technologies. The lack of these latter features make conventional holograms on a plate 
difficult to achieve practical usage for 3D streaming. To some extent, the recent 
multi-view LFD, lenticular-based auto-stereoscopic displays as the Looking Glass 
(LG) Portrait \cite{LG25}, can overcome these defects as we shall discuss below. 
The role of the lenticular lens is to transfer simultaneously the information of 
specific pixels of an image to some designated positions on a LCD screen. These are 
commercial devices and have the advantage to produce realistic 3D displays with a 
reasonable resolution without using additional external devices.

In the following, we present a new framework designed for the Linux environment,
and adapted to MS Windows OS, in order
to achieve practical 3D display in real-time by processing single 2D images or frames
and display these in a LG Portrait. The new algorithm does not require the use of heavy 
computing run-time or to wear specific glasses. As discussed, the visual quality of the 
synthesized 3D views in real-time can provide a rather realistic immersive experience.


\section{Architecture and implementation}
\label{sec:architecture}

\begin{figure}[H]
\includegraphics[width=14cm]{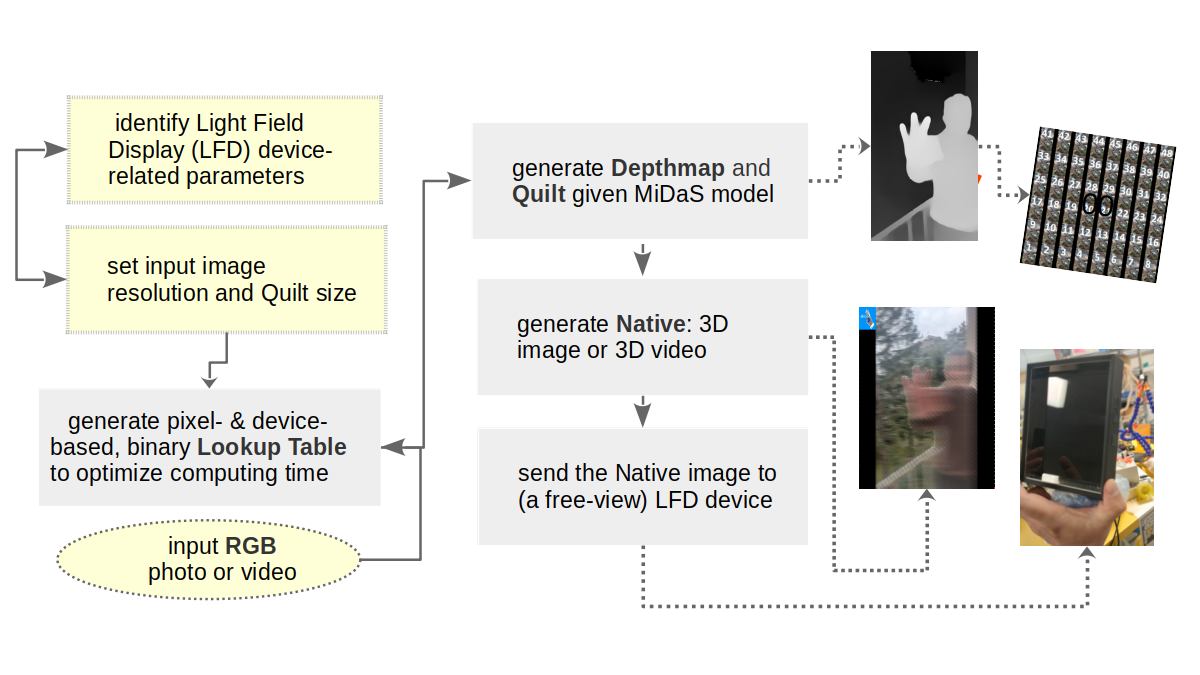}
\caption{Diagram of the processes involved in the altiro3D Library --for details see \cite{Ten24}. \label{fig:workflow}}
\end{figure}

The proposed system is a main improvement of our previous project named altiro3D. altiro3D is a highly specialized Library with many possibilities to directly 
feed holographic displays. The previous released toolkit version was able to create 3D views from a single image or from a video {\tt .mp4} file as illustrated in Fig.\ref{fig:workflow}.
The current version extends previous capabilities to real-time video output.

\unskip

\vspace{0.4cm}

The system is now able to acquire frames from a USB camera, to produce the views and to map the output directly in the Native display format. 
Now the altiro3D system can manage multiple displays: a new feature of the current system is the possibility to generate a video stream (\verb+altiro3Drt+, \verb+altiro3Dlive+ and \verb+altiro3Dlive-cli+) from a region chosen in the current screen (primary) of the PC and to directly send the processed result to the other display (secondary), connected to the external LFD.
Moreover the current version of \verb+altiro3Dlive+ provides a graphical user interface to easily create the mapping LUT from the Looking Glass Holoplay device, to configure and manage the acquisition and the generation of the stream. 

Another improvement of the current system is the possibility to run the application (and the related Library) under both Linux and Windows OS. All the used libraries have been properly chosen to be multi-platform. Moreover the adoption of the Docker containers (in order to simplify the compilation environment and to improve the repeatability of results) simplified the development loop. In particular we created a Docker container with a complete toolkit (Qt, OpenCV, FFmpeg and other libraries) to cross-compile and package the altiro3D system for Windows. 

In Section \ref{sec:workflow} a brief summary of the main processing steps is given from the 2D source (acquisition step) to the Native format (i.e. the format that the holographic display can directly show), then in Sec.~\ref{sec:implementation} the new developed tools for real-time acquisition and rendering are described. Then in Section \ref{sec:gui} the graphical user interface for screen recording and direct rendering is described in details. Moreover a brief description of the cross-compilation process and of the containerization approach to get a Windows installer is given.

\subsection{Workflow}
\label{sec:workflow}

The altiro3D system is able to create multiple views from a single image to feed an external LFD
as described in the workflow in Fig.\ref{fig:workflow}.
All tests so far have been done with the Holoplay Portrait Looking Glass device. However, the system and the provided Library, is device agnostic. Hence the code developed can be applied to any other free-view holographic display.

\begin{table}[h]
    \centering
    \begin{tabular}{c|ccc}
    \hline \hline
        {\bf Calibration}   & \multicolumn{3}{c}{Extraction of device configuration} \\
        \hline
        {\bf Acquisition}   & Real-time camera & Screen & Video or images \\
        {\bf MiDaS}         & \multicolumn{3}{c}{Depth map } \\
        {\bf View creation} & Geometric method & Fast method & \\
        {\bf Mapping}       & \multicolumn{3}{c}{Fast LUT mapping} \\
        {\bf Display}       & \multicolumn{2}{c}{Qt direct multiplatform display} & Save file\\
    \hline \hline
    \end{tabular}
    \caption{Workflow summary.}
    \label{tab:workflow}
\end{table}

The altiro3D system is now able to deal with real-time sources, such as stream from a USB camera (REAL) or stream acquired by ``snapshotting'' (as in the context of video games) from a region of screen (SNAP) connected to the PC.

While the acquisition from a USB camera (REAL) is not a big problem in both Linux and Windows (openCV provides a simple way to access camera devices), in order to produce a cross-platform toolkit it has been necessary to find a method to access the video screen data in both OSs. A public multi-platform project has been used \cite{screenrecorder} as a starting point for the screen acquisition (SNAP) and has been integrated in the altiro3D project. This project has been developed in C++ and it provides altiro3D the capability to get a video stream from a screen by using the FFmpeg Library API. With this improvement altiro3D is able to create a Native stream to directly feed an external LFD or to create a Native {\tt .mp4} video file.

As in \cite{Ten24} the altiro3D system exploits the MiDaS neural network to produce an accurate depth map of every acquired image.  The MiDaS network can run by using the CPU or CUDA: the network is called by the DNN (Depth Neural Network) of OpenCV which loads the ONNX format of the network. Some modifications of the original altiro3D source code provided a speed-up of the processing step.

As described in \cite{Ten24}, after the depth estimation, altiro3D can generate multi-views by using two algorithms (FAST and REAL) which can be chosen as a trade-off between accuracy and processing speed. 

In the next step the altiro3D code generates the Quilt image where all views are properly combined to render the holographic scene. As in the previous version \verb+altiro3Dlive+ exploits the LUT table to speed up the mapping from the Quilt representation to the Native final image. 
In order to get a better performance of the processing in real-time the \verb+parallel_for+ optimization of OpenCV has been employed: this construct provides a method to parallelize critical sections of the code (i.e., suitable for running on a parallel processing system), by exploiting the cores provided by the PC. The parallel\_for approach has been particularly used in the mapping process from the Quilt to the Native stage, where the problem that needs to be evaluated can be separated.

In order to provide the real-time video stream, a particular Qt implementation has been developed to reduce the processing cost and to provide a smooth video rendering.  

\subsection{Tools for 3D creation and development environment}
\label{sec:implementation}

As previously mentioned some new tools have been developed to provide real-time video stream for 3D displays.

\begin{itemize}
    \item altiro3Drt
    \item altiro3Dlive-cli
    \item altiro3Dlive (see Sec. \ref{sec:gui})
\end{itemize}

\verb+altiro3Drt+ provides the possibility to create a Native 3D video stream to be directly reproduced in the LFD starting from a video file or directly from a USB camera. This executable does not have a GUI and has to be configured and called from the command line. The parameters to be passed are:

\begin{itemize}
    \item path to the MiDaS network weights,
    \item path to the map LUT file,
    \item Quilt resolution,
    \item the Quilt mask according to the passed map file,
    \item optional sub-sampling factor to speed-up the processing,
    \item a configuration INI file, where a file or a USB camera index can be specified.
\end{itemize}

With this tool the user is able to feed a holographic display with a 2D video file (converted in 3D in real-time) or with a USB camera.

\verb+altiro3Dlive-cli+ allows to acquire from a local screen an image region and to convert the acquired 3D region to a 3D Native stream. The parameters to be passed to this executable are similar to those for \verb+altiro3Drt+. The main difference is in the configuration INI file: the user has to define the region of the screen to be acquired, the position inside the screen of these region. Moreover, in presence of multiple screens, the user has to specify the screen where the region is acquired and the screen where the Native stream is displayed. As mentioned, in order to find a multi-platform solution (for both Linux and Windows, and potentially also for MacOS) we adapted an open source code \cite{screenrecorder}: the code has been modified to fit in the altiro3D Library and some improvements of the code have been added. The ScreenRecorder class exploits multi threading:

\begin{itemize}
    \item a thread to acquire raw frames from the selected screen region,
    \item a thread to display the Native image to the output screen or another thread to save the Native encoded flames to the output {\tt .mp4} file.
\end{itemize}

The audio stream is not currently used in the altiro project.

\verb+altiro3Dlive+ is the GUI version of the previous executable and provides an easy interface to setup parameters, to start and stop the acquisition. It will be thoroughly described in Sec.~\ref{sec:gui}. The GUI version prevents the manual creation of the INI configuration file needed by \verb+altiro3Dlive-cli+.

As previously mentioned, the altiro3D project has been developed by taking into account the portability of the code. In order to optimize the cross-compilation process, containers have been used. 
The use of containers in the source code development is motivated by the diffusion of the continuous integration approach and by the possibility to create an insulated and repeatable environment both for testing (TDD, Test Driven Development) and compilation. All the necessary libraries for compilation or test can be integrated in an image container, ready to be re-started. The running container starts always from the same initial reproducible state. 

In order to provide a cross-compiled version of altiro a Docker environment \cite{docker} has been created to improve the repeatability of the compilation and to simplify the versioning. The Docker environment has been built by installing all necessary libraries to compile the code for Windows platform. The \verb+cmake+ compilation system used to compile the source code under Linux has been modified to be compiled with the mingw toolchain \cite{mingw}. This toolchain is able to produce a Windows executable which can be ran with some additional DLL's. Before compiling the altiro3D Library and tools, the Docker image compiles from scratch all necessary dependent libraries: Qt, OpenCV, FFmpeg and Hidapi. 
After compiling the mentioned libraries, some automatic scripts download the altiro source code, configure the project and compile the whole project.
After that another script is in charge to use the NSIS (Nullsoft Scriptable Install System \cite{nsis}) to create a Windows installer, where all dependent libraries are embedded.

It has to be noted that all these steps are automatically ran and do not need a user interaction.
The Docker image is based on Ubuntu 22.04 LTS and all cross platform libraries are properly installed when it is instantiated.

\subsection{Graphical user interface for screen recording altiro3Dlive}
\label{sec:gui}

In order to simplify and to complete the altiro3D toolkit a Graphical User Interface has been developed.
This GUI has been developed in Qt and provides a cross-platform solution which can be potentially compiled in Linux, Windows and MacOS. The main task of this GUI is to exploit the altiro3D Library while acquiring in SNAP mode the video stream, from the current user display. The converted video stream can be directly send to the external LFD or record a {\tt .mp4} file in Native format.

\begin{figure}[H]
\begin{adjustwidth}{-\extralength}{0cm}
\centering
\subfloat[\centering]{\includegraphics[width=8cm]{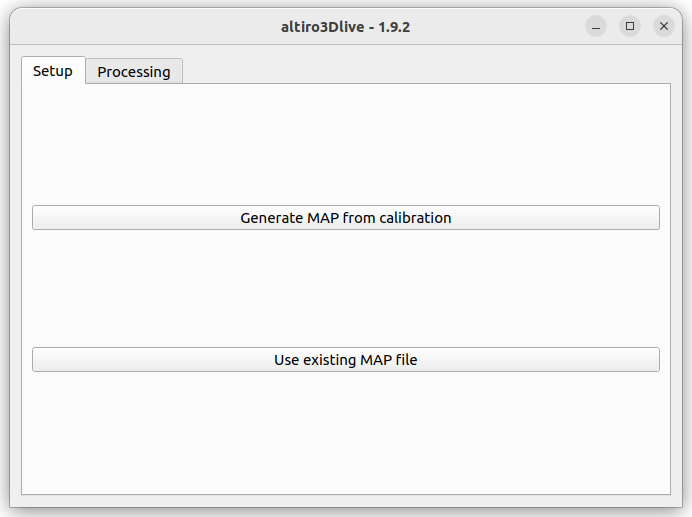}}
\subfloat[\centering]{\includegraphics[width=8cm]{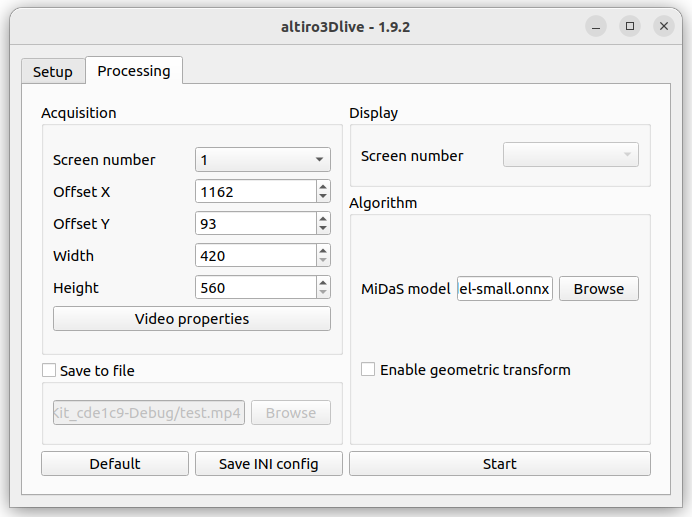}}\\
\end{adjustwidth}
\caption{Main GUI with two tabs: setup and processing.
(\textbf{a}) Setup tab.
(\textbf{b}) Processing tab.}
\label{fig:mainmenu}
\end{figure} 

The GUI provides users an easy method to setup (Fig.~\ref{fig:mainmenu}a) and to manage the real-time conversion (Fig.~\ref{fig:mainmenu}b).
In the setup tab Fig.~\ref{fig:mainmenu}a the user can choose between

\begin{itemize}
    \item \verb+Generate MAP from calibration+
    \item \verb+Use existing MAP file+
\end{itemize}

\begin{figure}[H]
\begin{adjustwidth}{-\extralength}{0cm}
\centering
\subfloat[\centering]{\includegraphics[width=6cm]{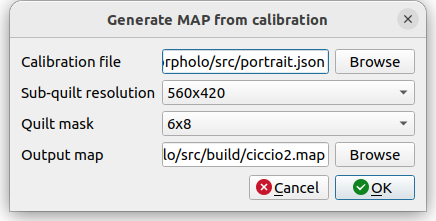}}
\subfloat[\centering]{\includegraphics[width=6cm]{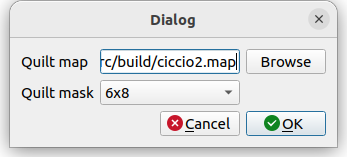}}
\subfloat[\centering]{\includegraphics[width=4.4cm]{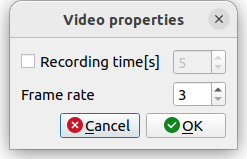}}\\
\end{adjustwidth}
\caption{Menus of the altiro3Dlive GUI.
(\textbf{a}) Generate MAP from calibration dialog. \label{fig:setup1dialog}
(\textbf{b}) Use existing MAP file dialog. \label{fig:setup2dialog}
(\textbf{c}) Video properties dialog. \label{fig:videoproperties}
}
\end{figure}

If the user selects the first choice, a dialog is shown (Fig.~\ref{fig:setup1dialog}a) which can be filled-in to create the LUT map file for further processing step. The user has to extract from the Holoplay Looking Glass the JSON file with the calibration file (copy and paste), then the optimum sub-Quilt resolution and the Quilt mask have to be selected, which is the number of views that will be generated by the altiro3D Library and that will be used by the holographic display. Finally the output MAP file has to be chosen. 

Instead, with the dialog in Fig.~\ref{fig:setup2dialog}b, the user can re-use an existing MAP file: the user has to choose the right Quilt mask (used when that MAP file has been generated). The system will use that file in the processing tab. 

\begin{figure}[H]
\begin{adjustwidth}{-\extralength}{0cm}
\centering
\subfloat[\centering]{\includegraphics[width=8cm]{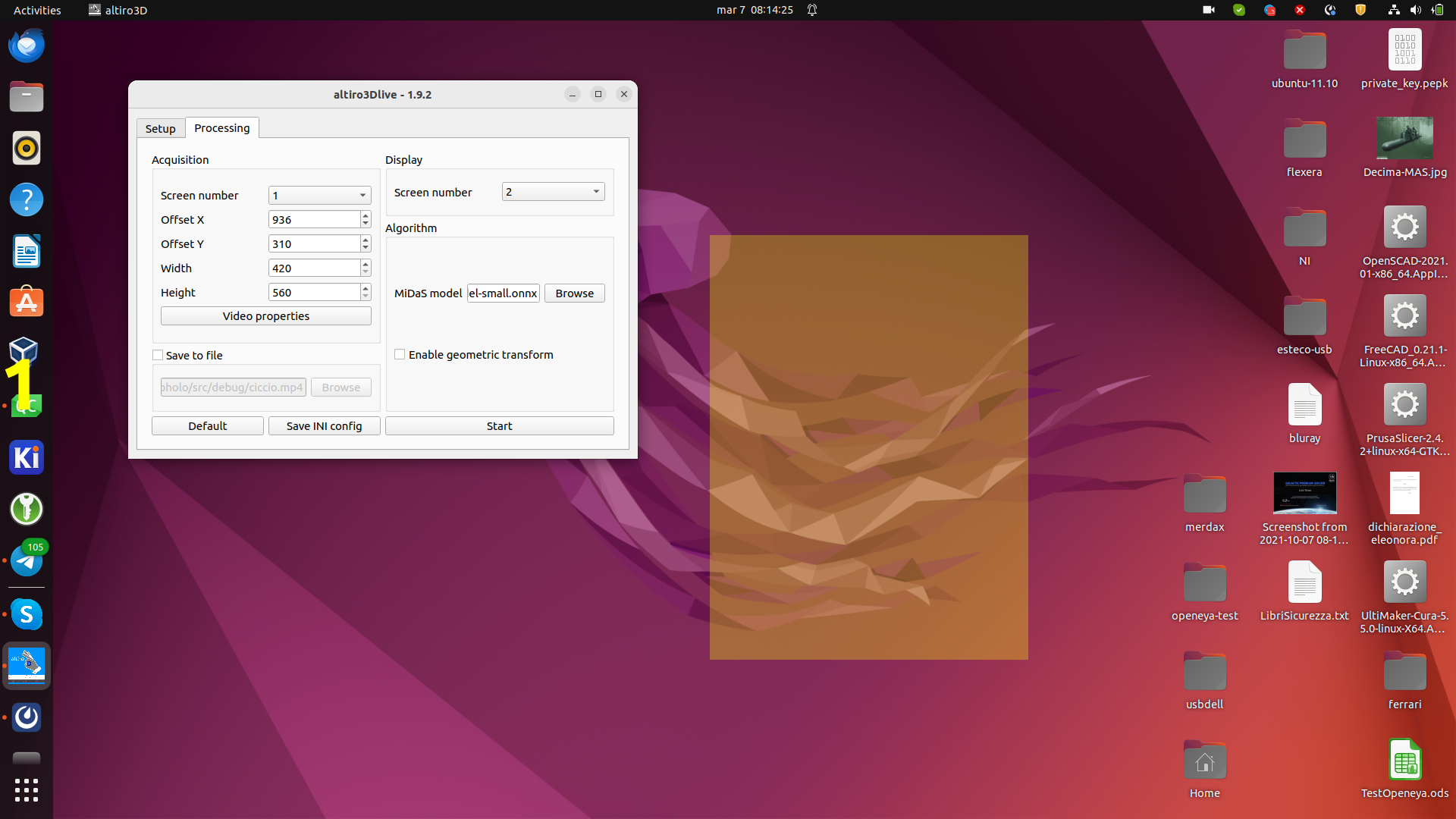}}
\subfloat[\centering]{\includegraphics[width=8cm]{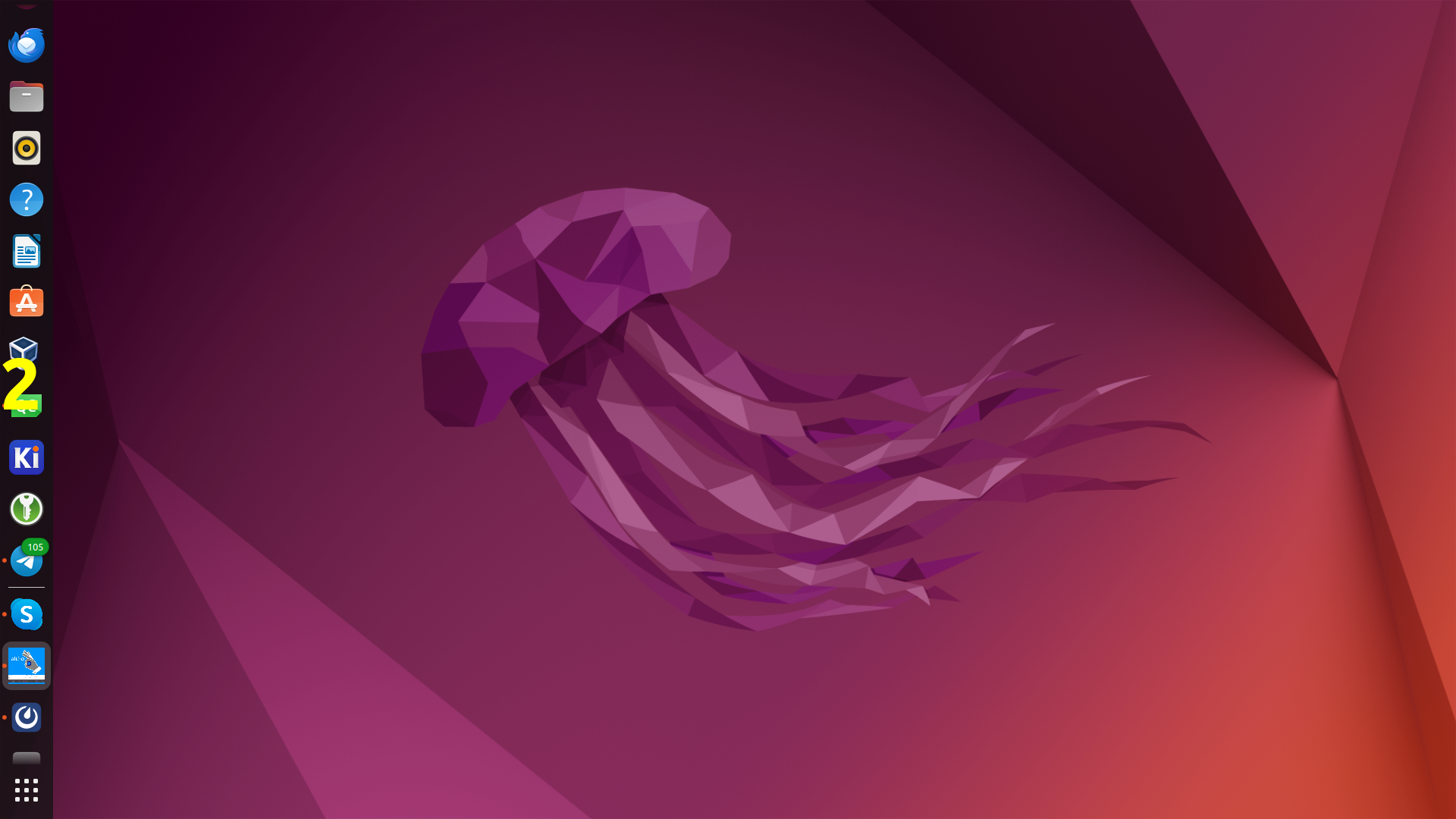}}\\
\end{adjustwidth}
\caption{Processing tab in action with multiple displays.
(\textbf{a}) Screen 1 and \label{fig:screen2}
(\textbf{b}) Screen 2 with altiro3Dlive. \label{fig:procinaction}
}
\end{figure}

The processing tab (Fig.~\ref{fig:mainmenu}b) is used to configure the acquisition and to start and stop acquisition. The processing tab detects the screens currently connected to the PC and shows the screen number in the left-middle region (see Fig.~\ref{fig:screen2}a and \ref{fig:screen2}b) with yellow color and shows the yellow region that will be acquired when acquisition starts (Fig.~\ref{fig:procinaction}).

The user can move the yellow region within the screen using drag and drop or he can manually setup the position, the size of the yellow region and the screen's number with the Acquisition sub-menu in the Processing tab. If the size of the region cannot fit the size for the Quilt creation (it mainly depends on the geometry of the target device), an adaptation process is used, which preserves the image ratio and eventually adds black side bars.
The Video properties button is placed in the Acquisition sub-menu. It opens another dialog (Fig.~\ref{fig:videoproperties}c) where the frame-rate and duration of the acquisition can be setup. If the recording time checkbox is not checked, the start and stop operation is manual.

If the Save to file groupbox is checked, the acquisition does not feed the output screen but saves the encoded Native video in the file path.

In the right side of the processing tab, it is possible to select the output display and some 
algorithm parameters if two displays are connected to the PC. With the sub-menu Display the output screen can be selected.
If the LFD is enumerated with number 2, the output display can be set to 2 to feed the holographic display with the Native real-time stream.
Moreover, in the Algorithm sub-menu the path to the MiDaS neural network file can be set and the geometric view creation algorithm can be enabled. If this is not enabled, the fast procedure is used.

In the bottom region the Start button can be used to start the acquisition. The application creates an icon tray which will be shown in the system bar of the OS. The user can right-click on the icon to manually stop the acquisition. Then the original windows is shown.
The new GUI simplifies all steps to create a real-time stream to feed an external LFD or to create a {\tt .mp4} file in a Native format.

\section{Results}
\label{sec:results}

\begin{figure}[H]
\begin{adjustwidth}{-\extralength}{0cm}
\centering
\subfloat[\centering]{\includegraphics[width=8cm]{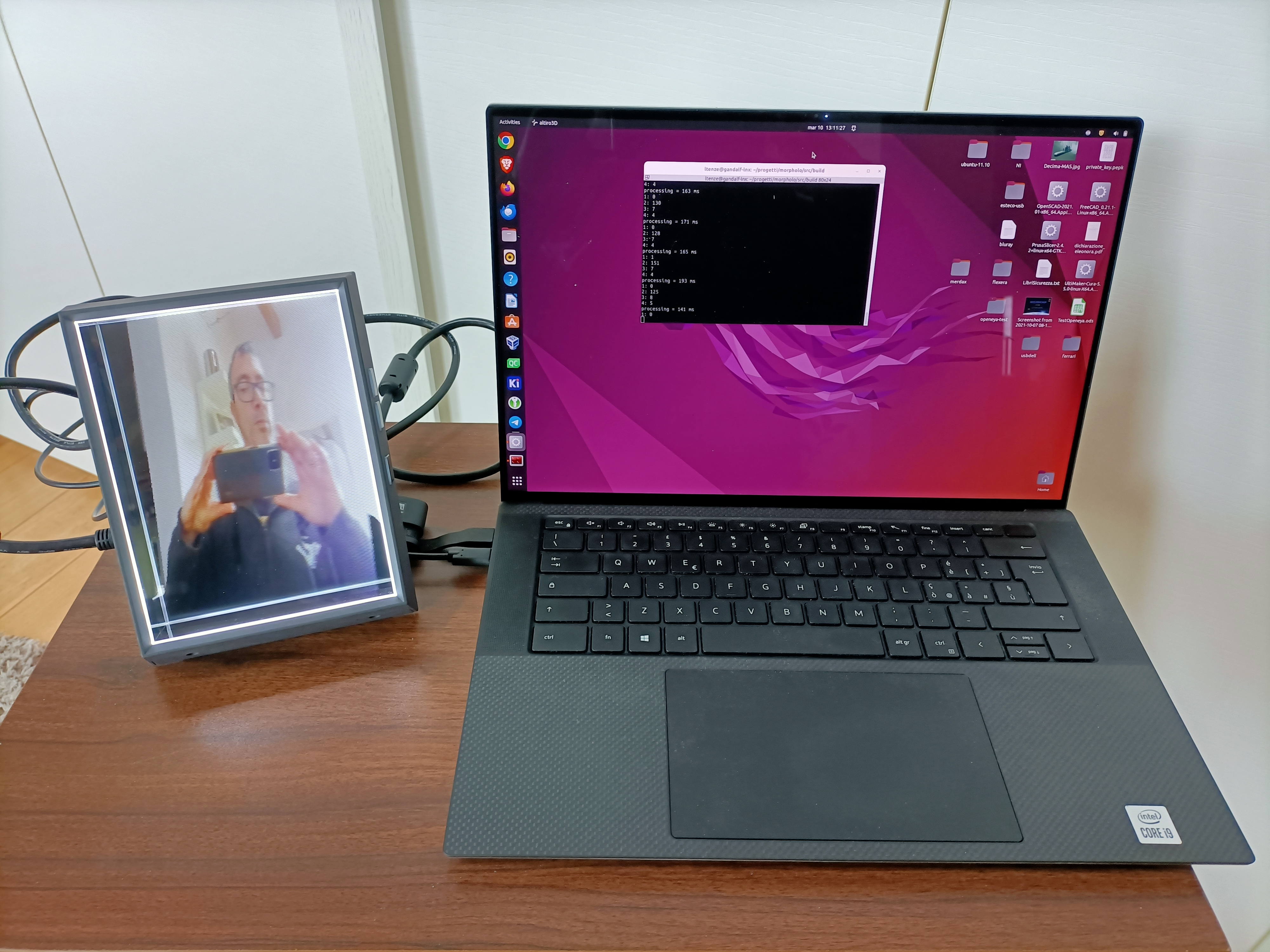}}
\subfloat[\centering]{\includegraphics[width=8cm]{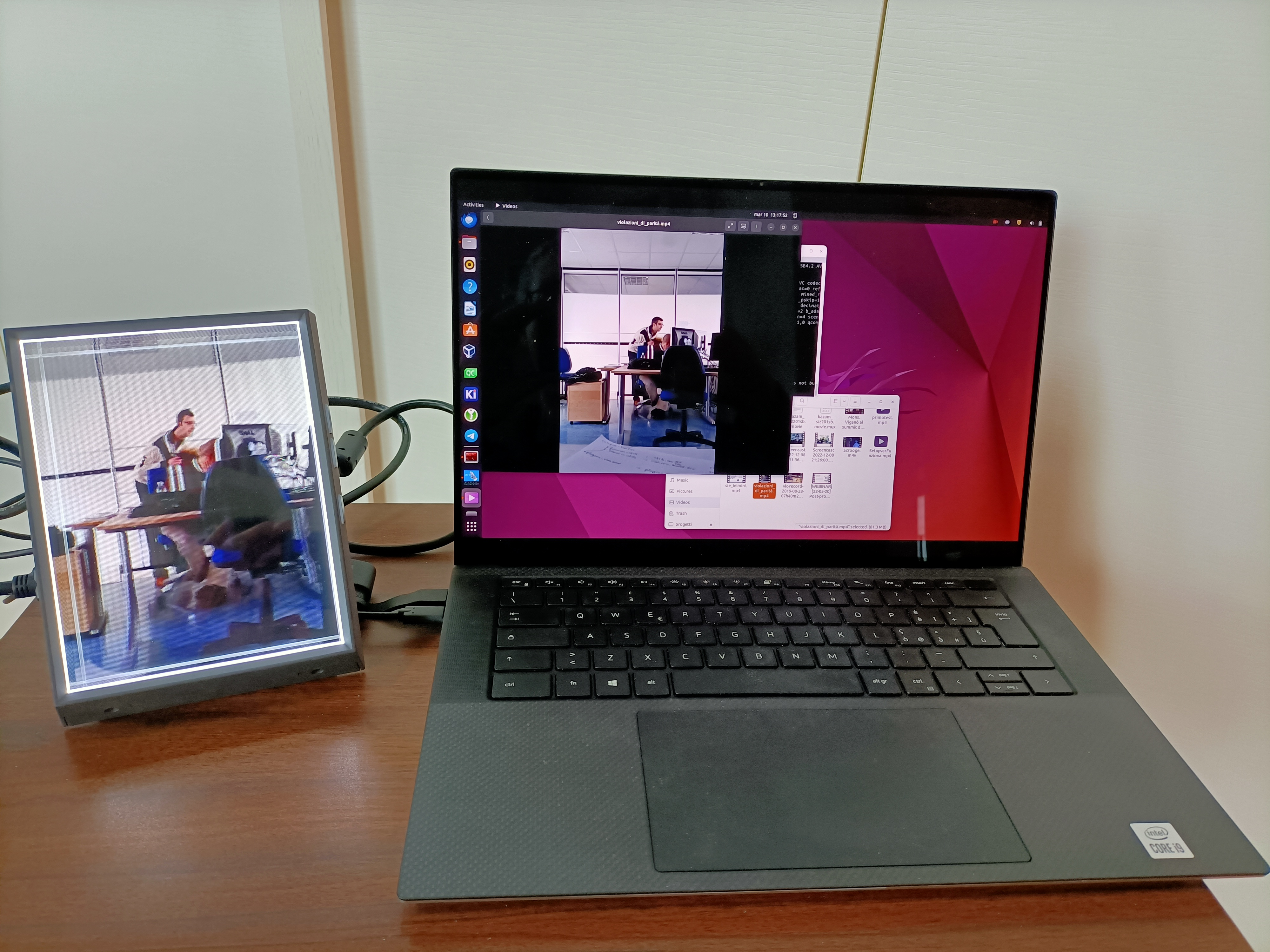}}\\
\end{adjustwidth}
\caption{REAL (left) and SNAP (right) acquisition mode.
(\textbf{a}) altiro3Drt in action: the stream from camera feeds the external holographic display. \label{fig:rt}
(\textbf{b}) altiro3Dlive in action: a region of the right screen is converted to the Native format and the real-time stream feeds the LFD. \label{fig:live}
}
    \label{fig:altiro_in_action}
\end{figure}

The current version of altiro3D has been tested to check the performance of the current implementation. 
In particular the altiro3Drt executable has been tested in a Dell Laptop Intel(R) Core(TM) i9-10885H CPU @ 2.40GHz with 16 cores equipped with a NVidia GeForce GTX 1650 Ti Mobile --as shown in Fig.~\ref{fig:altiro_in_action}

On this Laptop, under Linux, altiro3Drt can acquire frames from the embedded camera in real-time with a framerate of 10 Hz ($\thicksim$ 100 ms/frame) while exploiting a MiDaS small network version 2.1. It has to be noted that this performance has been reached without using the GPU's CUDA cores.

\begin{table}[b]
    \centering
    \begin{tabular}{ccccc}
    \hline \hline
         OpenCV & CPU & CUDA & CUDA FP16 & Processing time\\
         \hline
         4.5.4 & yes  & no & no & $\thicksim$100 ms/frame \\
         4.7.0 & yes  & yes & no & $\thicksim$160 ms/frame \\
         4.7.0 & yes  & no & yes & $\thicksim$100 ms/frame \\
    \hline \hline
    \end{tabular}
    \caption{Comparison of performance with CUDA with different OpenCV DNN target configuration. Decimation factor is 8 and the Quilt mask is $6\times8$}
    \label{tab:cuda}
\end{table}

When the OpenCV Library had been compiled with CUDA support and some tests have been performed to understand the behavior of the code using different configurations of the DNN OpenCV inference engine.
The CUDA back end can be used with two different targets configuration: CUDA or CUDA FP16. As shown in Table~\ref{tab:cuda} the behavior of the running code with CUDA target option performs worse than the CPU and CUDA FP16. The performance with CPU and CUDA FP16 option is almost the same. The OpenCV Library has been re-compiled with both CUDA and OpenVINO support. 

\begin{table}[h]
    \centering
    \begin{tabular}{ccccp{35ex}}
    \hline \hline
    Exclusive & Total & Inclusive & Total & Name\\
    CPU & & CPU \\
    seconds & \% & seconds & \% \\
    \hline
        8.166 & 100.00 &  8.166 & 100.00 &   <Total>\\
        2.652 &  32.48 &  2.652 &  32.48 &   <static>@0x28a8f2 \newline (<libopencv\_dnn.so.4.6.0>)\\
        1.771 &  21.69 &  4.513 &  55.27 &   <static>@0x14592b \newline  (<libopencv\_dnn.so.4.6.0>)\\
        0.881 &  10.78 &  1.131 &  13.85 &   Quilt2Native::Quilt2NativeParallel \newline
        ::operator()(cv::Range const\&) \\
        0.831 &  10.17 &  0.851 &  10.42 &   <static>@0x1a5292  \newline (<libopencv\_imgproc.so.4.6.0>)\\
        0.610 &   7.48 &  0.610 &   7.48 &   sched\_yield\\
        0.300 &   3.68 &  0.871 &  10.66 &   <static>@0x13870 (<libtbb.so.12.9>)\\
        0.240 &   2.94 &  0.240 &   2.94 &   QuiltMap::getMapPtr()\\
        0.190 &   2.33 &  1.051 &  12.87 &   <static>@0x19504d \newline (<libopencv\_imgproc.so.4.6.0>)\\
        0.170 &   2.08 &  0.170 &   2.08 &   <static>@0x16dc48 \newline (<libopencv\_dnn.so.4.6.0>)\\
        0.100 &   1.23 &  0.100 &   1.23 &   <static>@0x17cce1 \newline (<libc.so.6>)\\
    \hline \hline
    \end{tabular}
    \caption{Main results obtained by profiling altiro3Drt with gprofng}
    \label{tab:gprogng1}
\end{table}

In order to get a profiling of the code, the complete altiro3D suite has been compiled under  Ubuntu 24.04: altiro3D deeply exploits multi-thread programming and the new version of \verb+gprof+ (not suitable for multi-threading) tool is necessary. This new tool, \verb+gprofng+ (see \cite{gprofng}) is available from the updated binutils  package of Ubuntu, but it is present only after the 23.10 version. So we decided to profile the suite by using the Ubuntu 23.10 and new version 24.04 LTS. 

As it can be seen in Table~\ref{tab:gprogng1}, the most time consuming functions are those related to the MiDaS call: indeed the \verb+libOpenCV_dnn+ Library (2nd and 3rd line of the table) is the OpenCV Library which provides improved Deep Neural Network inference. The Library \verb+libtbb+ (4th line) is the one used by the OpenCV parallel\_for call to paralellize the evaluation. \verb+Quilt2NativeParallel+ (5th line) is the function implementing the parallelization with parallel\_for. \verb+QuiltMap::getMapPtr+ is the function used to implement the LUT.

From these results we can argue that the most expensive part is the MiDaS inference: as future work we could improve this evaluation by using OpenVINO's (see \cite{openvino}) framework to speed up the inference of the network. Another optimization we should consider is the improvement of the parallel\_for structure and a new improved implementation of steps from the view creation to the Native image: many memory movements can be reduced if the Quilt map creation is skipped and the Native image is directly mapped.

\section{Final Remarks}

We have developed a real-time stream algorithm based on a new extended version of the 
altiro3D C++ Library with AI-generated multiviews for the 3D (of any portion of a) 
Desktop display. This is achieved by selecting a Desktop region with the help of a 
simple GUI for the tagged screen acquisition area. The user is able to start 
any application within this 3D-capable GUI environment with minimal effort 
and by just one click. The current GUI can, at present, be only used for acquiring the video 
stream from the display. Further development can complete and improve the current 
GUI in order to include all functionalities of the Library and to simplify the user's 
interaction.

Our 3D-GUI, in its present form, opens the possibility of watching 2D YouTube 
videos or doing video conferencing in 3D without the need of installing any other 
specific, tailored, extra software for each application being in use. We discussed 
how the extended altiro3D Library performs satisfactorily on any standard PC running 
Linux O.S., without a 3D graphics accelerator board. 

As stated in Sec.~\ref{sec:results}, some improvement of the code can be taken into 
account to improve Library performance and speed-up real-time processing.
Our hope is that libraries such as altiro3D could inspire 
subsequent work on real-time 3D rendering with the use of AI and lenticular-based 
holographic screens by reducing exhaustive computation while endowing more accuracy 
and realistic 3D visions.

\vspace{6pt} 


\dataavailability{
Further information, binaries, papers, presentations, manuals, or to report bugs, can be found at
https://github.com/canessae/altiro3D/releases/tag/2.0.0
} 

\conflictsofinterest{The authors declare no conflicts of interest.} 


\reftitle{References}

\end{document}